\def\etal{{et al. }}

\def\tx{T_X}
\def\mvir{M_{\rm vir}}
\def\mvirtx{M_{\rm vir} {\rm -} T_X}
\def\rhocr{\rho_{\rm cr}}

\documentstyle[12pt,aaspp4,epsf]{article}
\begin{document}
\lefthead{Voit & Donahue}
\righthead{Temperature--Viral Mass Relation for Clusters}
\slugcomment{\apj  ~Letters, in press (submitted 1998 March 5)}
\title{On the Evolution of the Temperature--Viral Mass Relation
for Clusters of Galaxies}
\author{G. Mark Voit\footnote{voit@stsci.edu} 
     \& Megan Donahue\footnote{donahue@stsci.edu}} 
\affil{Space Telescope Science Institute \\3700 San Martin Drive \\
Baltimore, MD 21218 }

\begin{abstract}
X-ray temperature measurements of clusters of galaxies are now
reaching to redshifts high enough to constrain $\Omega_0$.
A redshift-dependent relation that maps these X-ray temperatures to 
the virial masses of clusters is an essential ingredient when one is
trying to determine cosmological parameters from cluster evolution.
Most such relations assume that clusters form from top-hat perturbations
that virialized just before the time we are observing them.  The smaller
$\Omega_0$ is, the less accurate these relations become.  Here we derive
a relation between virial mass and cluster temperature that allows for 
the fact that clusters form gradually and cease forming when the density
of the universe drops well below the critical value.  We show how 
sensitively the expected redshift distribution of clusters depends on
the mass-temperature relation used and argue that one needs to use
a relation that yields no evolution in the low-$\Omega_0$ limit.
\end{abstract}


\section{Introduction}

Studies of temperature evolution in massive clusters of galaxies
are now realizing their promise as cosmological indicators.  
The most massive clusters in the
universe mark the rarest, highest amplitude peaks in the intial 
perturbation spectrum that seeded structure formation and
provide powerful leverage on parameters such as $\Omega_0$, the 
current matter density of the universe in units of the critical 
density $\rhocr$.  If $\Omega_0 = 1$, the development of structure 
should have proceeded unabated from early in time to the present day. 
However, the relative lack of evolution observed in the X-ray 
temperature function 
of clusters indicates that the formation of structure is stalling, 
and thus that $\Omega_0 \approx 0.2-0.5$ (Henry 1997;
Bahcall \etal 1997; Carlberg \etal 1997; Donahue \etal 1998)

Using X-ray observations of clusters to constrain cosmological
models requires knowledge of how the X-ray temperature of a cluster 
($\tx$) relates to its mass.  Temperatures of distant clusters 
are now straightforward to measure (e.g., Donahue 1997; 
Donahue \etal 1998; Henry 1997),
but relating these temperatures to cluster masses is a subtler art.
In order to define a cluster's mass, one must decide upon an 
appropriate boundary separating the cluster itself from peripheral 
material not yet part of the cluster.  
Theorists tend to choose a boundary that encompasses matter 
that has virialized and excludes matter that is still falling into the
cluster.  The mass within this boundary is then $\mvir$, the virial 
mass of the cluster.

Defining a cluster's mass to be $\mvir$ has an established theoretical
pedigree, but this definition is observationally problematic.  
Virial masses are a staple of the Press-Schechter approach to the 
problem of hierarchical structure formation, which has become enormously 
popular owing to its basic agreement with much more sophisticated 
$n$-body techniques (e.g., Lacey \& Cole 1994).  
Press \& Schechter (1974) presumed 
that the collapse of a spherical top-hat perturbation would adequately 
represent the formation of more complicated objects.  In such a 
spherically symmetric collapse, the density of a newly formed object 
just after collapse is $18 \pi^2 \rhocr \approx 178 \rhocr$ in an 
$\Omega_0 = 1$ universe, but this contrast factor drops to as low as 
$8 \pi^2 \rhocr \approx 79 \rhocr$ in an open universe with 
$\Omega_0 \ll 1$ (e.g., Lacey \& Cole 1993).  Even if one could
measure such density contrasts accurately, {\em a priori} knowledge 
of $\Omega_0$ would still be needed to determine $\mvir$. Thus, the 
bridge between $\tx$ and $\mvir$ must be built primarily from the 
theoretical side.

Numerical models of cluster formation show that cluster temperatures
should indeed be closely related to their masses (e.g., Evrard \etal 1996).
To lowest order, cluster potentials that arise in such models are 
isothermal, with $T_X \propto M(r) / r$ (e.g., Cole \& Lacey 1996).  
If the mean density within radius $r$ is $\Delta$ times the
critical density, then the mass within that radius is 
$M(\Delta) \propto T_X^{3/2} \rhocr^{-1/2} \Delta^{-1/2}$.  
Proceeding in the vein of the Press-Schechter approach, one can 
define $\Delta_{\rm vir}$ to be the density contrast of a spherical 
top-hat perturbation just after collapse and virialization, in which 
case $\mvir = M(\Delta_{\rm vir})$ and $T_X \propto \mvir^{2/3} 
\rhocr^{1/3} \Delta_{\rm vir}^{1/3}$.  The density contrast parameter $\Delta_{\rm vir}$ depends in general on $\Omega_0$, the redshift 
of virialization, and the value of the cosmological constant 
(e.g., Eke \etal 1996; Oukbir \& Blanchard 1997).
Adopting the recent-formation approximation, which assumes 
that the clusters we now observe virialized just before their
light departed, then leads to a relation like the following:
\begin{equation}
\label{ecf}
  k T_X = (1.38 \, {\rm keV}) \beta^{-1} \left( \frac {\mvir} {10^{15}
      h^{-1} M_\odot} \right)^{2/3} \left[ \frac {\Omega_0}
      {\Omega(z)} \right]^{1/3} \Delta_{\rm vir}^{1/3} (1+z) \; \; ,
\end{equation}
where $\Omega(z)$ is the matter density of the universe in units of
$\rhocr$ at redshift $z$, $\beta$ is the usual ratio of potential depth
to X-ray temperature, and $h$ is the Hubble constant in units of
$100 \, {\rm km \, s^{-1} \, Mpc^{-1}}$ (e.g., Eke \etal 1996). 
In the limit of $\Omega(z) \approx 1$, we arrive at the scaling 
relation $T_X \propto \mvir^{2/3} (1+z)$.  

The shortcomings of this approach are well known (e.g. Viana \& Liddle 
1996; Kitayama \& Suto 1996; Eke \etal 1996).  Real clusters do not form 
at a single moment but rather accrete their matter over a long period of 
time.  If the accretion rate remains sufficiently high, as would be the 
case if $\Omega_0 = 1$, then the clusters we now see attained their 
observed masses relatively recently, validating the recent-formation
approximation that led to equation (\ref{ecf}).  If instead $\Omega_0 \ll 1$, 
then the recent formation approximation grows increasingly less valid 
in time.  Note that in the limit $\Omega_0 \rightarrow 0$, cluster 
evolution ceases by $z = 0$ and the $\mvirtx$ relation should no longer 
depend on redshift, whereas equation (\ref{ecf}) dictates $T_X \propto 
M_{\rm vir}^{2/3} (1+z)^{2/3}$.  One should also keep in mind that the
temperature of a real cluster depends on formation epoch only insofar as
the formation epoch affects the shape of the halo density profile and 
furthermore that real clusters are not necessarily isothermal.  
Comparisons of simulated clusters with equation (\ref{ecf}) indicate
that the effective value of $\beta$ declines modestly but systematically 
with time in an open universe with $\Omega_0 = 0.3$, slightly diminishing 
the expected amount of temperature evolution (Eke, Navarro, \& Frenk 1998).

While the relation in equation (\ref{ecf}) is adequate for many purposes, 
one would prefer to use a relation valid in the low-$\Omega_0$ limit
when constraining $\Omega_0$ through cluster temperature evolution. 
One way to patch the recent-formation approximation is to define an 
explicit formation redshift $z_f$ at which a cluster virializes.  
One can then integrate over the appropriate distribution of formation 
redshifts to determine the properties of observed clusters at 
redshift $z$ (Kitayama \& Suto 1996; Viana \& Liddle 1996).  

Motivated by the fact that clusters form gradually, not instantaneously, 
we present here an alternative approach to characterizing the redshift 
evolution of the $\mvirtx$ relation which has the advantage of being
qualitatively correct in the low-$\Omega_0$ limit.  In \S~2 we idealize
clusters as forming from spherically symmetric perturbations with 
negative radial density gradients rather than from top-hat perturbations.
This approximation alleviates the need to specify an artificial formation
redshift because it explicitly accounts for how clusters grow with time.
It also yields an analytical formula for the change in virial energy 
with virial mass.  We then use the merging-halo formalism of Lacey \&
Cole (1993) to derive an expression for the time-averaged mass accretion 
rate of a massive cluster, which allows us to determine how the $\mvirtx$ 
relation should change with time.  Section~3 briefly illustrates 
some implications of this relation. 

\section{Mass-Temperature Relation in an Open Universe}

Here we derive in three steps how the $\mvirtx$ relation should evolve.
First we analyze how a radially stratified, spherically symmetric 
perturbation would virialize.  Then we determine how the mass accretion
rate of a massive cluster changes with time in a Gaussian density field
with a power-law perturbation spectrum.  Combining the results of these
analyses, we determine how the $\mvirtx$ relation evolves with time, and
we normalize this relation using equation (\ref{ecf}).  For brevity and 
clarity, we restrict our derivation to the analytically tractable case 
of an open universe without a cosmological constant.

\subsection{Spherical Collapse with a Density Gradient}

Let us idealize the perturbation that will form a cluster as having 
a spherically symmetric density profile that declines with radius $r$.  
We then can define the Lagrangian coordinate $M$ to be the mass inside 
a given spherical shell centered on the origin. The evolution of each
shell depends only upon the mass inside of it.  In the absence
of a cosmological constant, these shells will obey the familar
parametric solution $r(M) = [GM(t_M/2 \pi)^2]^{1/3} (1 - \cos \theta_M)$, 
$t = (t_M / 2 \pi) (\theta_M - \sin \theta_M)$, where $\theta_M$
parametrizes the evolution of the shell containing mass $M$, which 
formally recollapses to the origin at time $t_M$ (e.g., Peebles
1993).

Drawing an analogy between top-hat collapse and the collapse of
concentric shells, we can approximate the virial mass of the
forming cluster to be the $M$ of the shell that has just
recollapsed to the origin.  The energy this infalling shell 
contributes to the total is equal to its potential energy at 
maximum expansion.  Because the maximum radius of the shell that 
converges upon the origin at time $t$ is $2[GM(t/2 \pi)^2]^{1/3}$, 
the virial energy $-E$ of the cluster accumulates according to
\begin{equation}
\label{dedm}
  \frac {dE} {d\mvir} =  \frac {1} {2} \left( \frac {2 \pi G \mvir} {t}
                         \right)^{2/3} \; \; .
\end{equation}
An expression for $\mvir(t)$ then provides the information needed
to determine $E(t)$ and $T(t) \propto E(t)/\mvir(t)$.

\subsection{Mass Accretion Rate for Massive Clusters}

The merging-halo formalism of Lacey \& Cole (1993) can be used to
derive a mean mass accretion rate for massive clusters.  This approach
extends the Press-Schechter picture by considering how clusters grow
via accretion of smaller virialized objects.  Mass accretion becomes
quasi-continuous for very massive clusters, and in this limit the Lacey
\& Cole formalism yields a mass accretion rate that enables us
to integrate equation \ref{dedm}.

In order to analyze halo merger rates, one defines 
$\delta({\bf x},t;M)$ to be the local fractional overdensity of the
universe, smoothed on mass scale $M$ and centered on comoving point 
${\bf x}$ at time $t$. Early in time the perturbations described by 
$\delta$ are assumed to be Gaussian, and in the linear regime they 
grow in proportion to the function $D(t)$, which depends on $\Omega_0$.  
While these perturbations remain linear, their rms amplitude on scale $M$ 
can be expressed as $\sigma(M) D(t) / D(t_0)$.  Ultimately some of them 
grow non-linear, and they are assumed to viralize when their amplitudes, 
extrapolated from the linear regime according to $D(t)$, exceed some 
critical threshold $\delta_c(t)$.  One can then trace the merger history
of a mass parcel beginning at ${\bf x}$ from time $t_1$ to the present
by keeping track of the largest $M$ for which $\delta({\bf x},t_1;M) D(t)
/ D(t_1) > \delta_c(t)$.  This largest mass is the mass of the virialized
halo containing the mass parcel at time $t$, and it jumps by an amount
$\Delta M$ each time the halo merges with another of mass $\Delta M$.

Equivalently, one can analyze the growth of structure by keeping the
amplitudes of the perturbations fixed, with rms amplitude $\sigma(M)$,
and tracking when they exceed the steadily falling threshold $\omega(t) 
\equiv \delta_c(t) D(t_0) / D(t)$.  In an open universe with no cosmological
constant, Lacey \& Cole (1993) show that 
\begin{equation}
  \omega(t) = \frac {3} {2} D(t_0) [ 1 + (t_\Omega/t)^{2/3}]
\end{equation}
where $t_\Omega = (\pi \Omega_0 / H_0)(1-\Omega_0)^{-3/2}$.  The parameter
$\omega$ thus serves as a useful surrogate for the time coordinate in
an open universe because it asymptotically approaches a fixed value as
growth of structure slows to a halt.

Now we are ready to determine an expression for the halo accretion rate.
Lacey \& Cole (1993) show that the probability that the mass of a virialized halo will jump from $M_1$ into an interval $dM_2$ at $M_2$ within a time interval corresponding to $d\omega$ is
\begin{equation}
  \frac {d^2p} {dS_2 d\omega} (S_1 \rightarrow S_2 | \omega) dS_2 d\omega
        = \frac {1} {(2 \pi)^{1/2}} 
          \left[ \frac {S_1} {S_2(S_1-S_2)} \right]^{3/2} 
          \exp \left[ - \frac {\omega^2 (S_1 - S_2)} {2S_1S_2} \right]
          dS_2 d\omega \; \; ,
\end{equation}
where $S_i \equiv \sigma^2(M_i)$.  To find the mean growth rate of a halo 
of mass $M$, we can integrate over this probability distribution as follows:
\begin{equation}
  \left\langle \frac {dS} {d\omega} \right\rangle =
      \int_{S_2}^\infty (S_1 - S_2) 
       \frac {d^2p} {dS_2 d\omega} dS_1 \; \; .
\end{equation}
The variance $S(M)$ generally declines with $M$, so for a rare high-mass
object, $S_2(M_2) \ll \omega^2$.  In this limit, the integrand's exponential
behavior restricts its main contribution to a narrow range where 
$S_1 \approx S_2$, reflecting the fact that most of the accreted mass 
comes from objects much smaller than the cluster. Thus,
\begin{equation}
  \left\langle \frac {dS} {d\omega} \right\rangle \approx 
      \frac {S} {\omega} \; \; .
\end{equation}
If the fluctuation amplitudes obey $\sigma(M) \propto M^{-(n+3)/6}$, where
$n$ is the usual power-law perturbation index in wavenumber space, we 
finally arrive at the mass accretion law $\mvir \propto \omega^{-3/(n+3)}$.
This result is equivalent to Lacey \& Cole's (1993) finding that 
$\mvir \propto t^{2/(n+3)}$ for $\Omega_0 = 1$.

\subsection{Evolution and Normalization of $\mvirtx$}

The typical temperature of a cluster should be proportional to the
virial energy divided by $\mvir$.  If we define $x \equiv 
[1 + (t_\Omega/t)^{2/3}]$ and $m \equiv 5/(n+3)$, we can write $\mvir 
\propto x^{-3m/5}$ and $dE \propto (x-1) dx^{-m}$.  Integrating 
over $x$ then gives $E \propto \mvir^{5/3} (x - 1 + 1/m)$. Thus,
we obtain
\begin{equation}
  T_X \propto \frac {E} {\mvir} \propto \mvir^{2/3}
              \left[ \left( \frac {t_\Omega} {t} \right)^{2/3}
                  + \left( \frac  {n+3} {5} \right) \right] \; \; .
\label{vdrel}
\end{equation}
This relation behaves properly in all the appropriate limits.
The time-dependent factor varies like $(1+z)$ when
$\Omega(z) \approx 1$, like $\rho_{cr}^{1/3} \propto t^{-2/3}$
for intermediate $\Omega(z)$, and goes to a constant 
as $\Omega(z) \rightarrow 0$.  To normalize this expression, 
we can simply match it to equation (\ref{ecf}) at early times:
\begin{equation}
 k T_X = (2.76 \, {\rm keV}) \beta^{-1} \left( \frac {\mvir} {10^{15}
      h^{-1} M_\odot} \right)^{2/3} \frac {1-\Omega_0} {\Omega_0^{2/3}}
              \left[ \left( \frac {t_\Omega} {t} \right)^{2/3}
                  + \left( \frac {n+3} {5} \right) \right] \; \; .
\end{equation}
Reducing $\Omega_0$ at fixed $\mvir$ and $t$ pushes $T_X$ upward 
because the effective formation epoch moves to earlier times, 
when the universe was denser.
Note also that the implied asymptotic behavior of temperature evolution 
depends on the power-law index $n$.  Larger amounts of power on small 
scales (larger $n$) lead to a less sensitive dependence of $\mvir$ 
on $\omega$ and an earlier cutoff for the accumulation of mass.

The true test of a relation between $\mvir$ and $\tx$ is how 
accurately it reproduces the evolution of the cluster temperature 
function in a large numerical simulation.  Showing that $M(\Delta)$ 
correlates well with $T_X$ for some chosen $\Delta$ is not sufficient 
because such tests do not probe how $\Delta_{\rm vir}$ changes 
with time.  As long as clusters are nearly isothermal above the 
density contrast $\Delta$, one will find $T_X \propto [M(\Delta)]^{2/3} 
\rho_{\rm cr}^{1/3} \Delta^{1/3}$.  If the goal is to constrain
cosmological parameters by applying Press-Schechter analyses to
observed clusters, then the $\mvirtx$ relation one uses should 
yield cluster temperature functions that approximate the results 
of numerical simulations as closely as possible.
To accurately reproduce behavior at low $\Omega$, one needs a 
relation like equation (\ref{vdrel}) that asymptotically yields 
no evolution.  In the most general cases where a single value of 
$n$ does not adequately represent the perturbation spectrum, 
equation (\ref{vdrel}) should still be useful as a fitting formula 
in which the asymptotic constant is a free parameter.

\section{Press-Schechter Predictions and $\mvirtx$ Evolution}

The generally good agreement between analytical Press-Schechter models 
of cluster evolution and numerical simulations has inspired numerous
applications of the Press-Schechter approach to X-ray cluster surveys
(e.g., Oukbir \& Blanchard 1997; Henry 1997; Carlberg \etal 1997;
Donahue \etal 1998).  
An accurate $\mvirtx$ relation is a crucial element in such
studies because it maps the virial masses used in the Press-Schechter
formalism to either observed X-ray temperatures or observed X-ray 
luminosities ($L_X$) through an $L_X {\rm -} \tx$ relation.  Here 
we will show that the consequences of using an inaccurate $\mvirtx$ 
relation can be quite significant.

For the purposes of studying cluster evolution, the change in 
$\mvirtx$ with $z$ is more important than its normalization.
Figure~1 shows how the quantity $\tx(\mvir,z)/\tx(\mvir,0)$
changes for four different relations: the one derived here for
$n = -1$ and $-2$, the relation from equation~(\ref{ecf}), and the
relation $\tx \propto \mvir^{2/3} (1+z)$ that one obtains by assuming 
that the mean density within the virial radius is some
constant multiple of the background density.  We have also computed
the mean temperatures for clusters of a fixed mass at a variety
of redshifts by integrating over formation redshifts following
Viana \& Liddle's (1996) method and with Lacey \& Cole's (1993)
definition of formation epoch.  The points on Figure~1 show how the mean
temperatures of these clusters that are originally 3$\sigma$ 
perturbations evolve for $n = -1$.  Note the close agreement between 
these points and the $n = -1$ $\mvirtx$ relation from
the present paper.

The differences between the $\mvirtx$ relations in Figure~1 are relatively 
modest in percentage terms, but they become greatly amplified when
filtered through the exponential distribution of cluster masses
that emerges from Press-Schechter calculations.  According to such
calculations, the differential comoving number density of clusters 
$dn$ in temperature interval $d\tx$ is given by
\begin{equation}
  \frac {dn} {d\tx} = \left( \frac {2} {\pi} \right)^{1/2} 
			\frac {3} {2}
			\frac {\Omega_0 \rho_{\rm cr}(0)} 
		              {T_X \mvir(T_X,z)}
			\frac {d \ln \sigma} {d \ln \mvir} \,
			\nu_c(\tx,z) \, \exp[ -\nu_c^2(\tx,z)/2 ] \; \; .
\label{tps}
\end{equation}
For a power-law perturbation index $n$ and a fiducial temperature
$T_0$, one can write $\nu_c(\tx,z) = \nu_{c0} [\omega(z)/\omega(0)] [\mvir(\tx,z)/\mvir(T_0)]^{(n+3)/6} $, where $\nu_{c0}$ or some analogous
parameter is adjusted to fit low-redshift cluster surveys: fixing 
$\nu_{c0}$ is equivalent to fixing the amplitude of the power spectrum
for a given $n$.  The derived number density of clusters thus depends 
exponentially on $\mvir(\tx,z)$.

An $\mvirtx$ relation that overestimates temperature evolution
will underestimate the virial mass corresponding to a high-$z$
cluster of a given temperature or luminosity.  Because the number 
density of clusters above a given mass decreases with virial mass, 
an underestimate of virial mass leads one to {\em overestimate} 
the numbers of high-$z$ clusters that ought to be in a particular
temperature-limited or flux-limited sample.  This effect is more 
severe in a flux-limited sample because the expected luminosity 
is much more sensitive to changes in the virial mass. 

To illustrate these effects, we have computed the surface density 
of clusters on the sky given by equation~(\ref{tps}) for $\Omega_0
= 0.2$, $n = -1$, various $\mvirtx$ relations, the $L_X {\rm -} T_X$
relation from Edge \& Stewart (1991), which we assume remains constant
with redshift (e.g., Mushotzky \& Scharf 1997), and a particular set 
of selection criteria.  We restrict the clusters to have $T_X > 
5 \, {\rm keV}$ and $F_X > 10^{13} \, {\rm erg \, cm^{-3} \, s^{-1}}$ 
within a $2.4^\prime \times 2.4^\prime$ detection cell like that 
of the Einstein Extended Medium Sensitivity Survey (EMSS; Henry 
\etal 1992).  We also include the effects of 
Galactic soft X-ray absorption, assuming a mean hydrogen column 
density $3 \times 10^{20} \, {\rm cm}^{-2}$.
Figure~2 shows the resulting surface densities, binned in redshift
intervals of 0.1.  Note that that at $z \sim 0.5$ the expected
cluster counts differ by a factor $\sim 3$, with more modest
temperature evolution yielding fewer expected clusters.
This factor is similar to the discrepancy Oukbir \& Blanchard (1997)
find between the actual redshift distribution of clusters in the
EMSS and their expectations, suggesting that the entire discrepancy
might stem from the $\mvirtx$ relation they use.  Above $z \sim 0.5$,
the flux limit becomes more restrictive than the temperature limit, and
the discrepancies between the expected cluster counts grow more pronounced.

Clearly, Press-Schechter approaches to modelling X-ray surveys 
require an accurate $\mvirtx$ relation.  The relation we give here
is likely to describe temperature evolution more accurately at late 
times than those based on the recent-formation approximation 
because it correctly yields no evolution as $\Omega(z) \rightarrow 
0$.  We are currently using this $\mvirtx$ relation to study 
cluster evolution in the EMSS and will report our results in 
a future paper.

\acknowledgements 

We would like to thank Caleb Scharf for his helpful suggestions
while we were developing these ideas.  We acknowledge NASA grants
grant NAG5-3257 and NAG5-3208 for partial support.


\newpage

\begin{figure}
\plotone{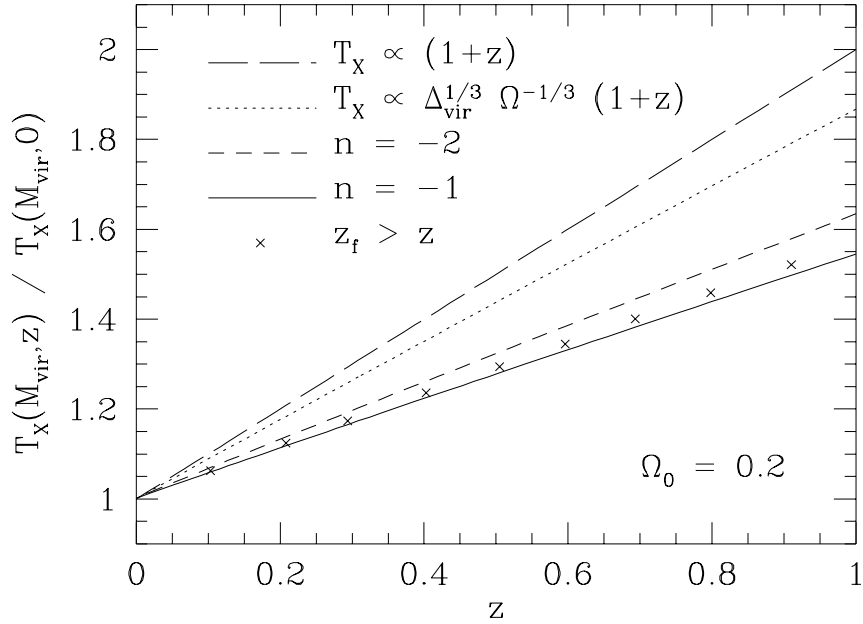}
\figcaption[mtdev.ps]{Temperature evolution for various $\mvirtx$
relations. The naive relation in which $\tx \propto \mvir^{2/3} (1+z)$ 
(long-dashed line) predicts the most evolution in $\tx$ for a given 
$\mvir$. Temperature evolution is more modest under the late-formation approximation in an open universe (dotted line) and is milder still 
for the relation derived in equation (\ref{vdrel}), given $n = -1$
(solid line) and $n = -2$ (short-dashed line). The points indicate
how mean cluster temperatures evolve in models that integrate over 
a range of formation redshifts $z_f$ to find the temperatures of 
clusters at a given redshift $z$, given an $n = -1$ perturbation
spectrum.} 
{\label{mtdev}}
\end{figure}

\begin{figure}
\plotone{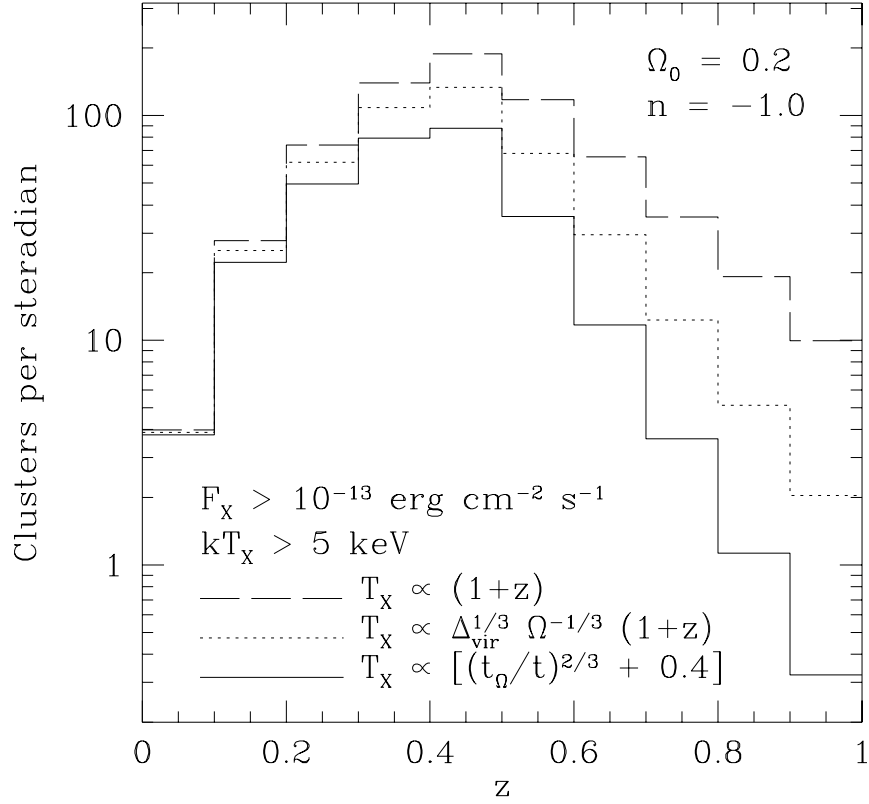}
\figcaption[nclust5.ps]{Redshift distribution of clusters 
as predicted by different $\mvirtx$ relations.  Here we show
expectations for clusters hotter than 5~keV with fluxes exceeding
$10^{13} \, {\rm erg \, cm^{-2} \, s^{-1}}$ in the EMSS detect cell.
The naive relation $\tx \propto \mvir^{2/3} (1+z)$ (dashed line) 
overpredicts the high-$z$ clusters because it overestimates 
temperature evolution and maps the flux and temperature limits 
to lower values of $\mvir$ than it should.}
{\label{nclust1}}
\end{figure}

\end{document}